 \documentclass[acmsmall,screen,authorversion,nonacm]{acmart}
\AtBeginDocument{%
  }

\copyrightyear{2026}
\acmYear{2026}
\setcopyright{cc}
\setcctype{by}
\acmConference[FAccT '26]{The 2026 ACM Conference on Fairness, Accountability, and Transparency}{June 25--28, 2026}{Montreal, QC, Canada}
\acmBooktitle{The 2026 ACM Conference on Fairness, Accountability, and Transparency (FAccT '26), June 25--28, 2026, Montreal, QC, Canada}
\acmDOI{10.1145/3805689.3806538}
\acmISBN{979-8-4007-2596-8/2026/06}

\usepackage{cleveref}

\usepackage[T1]{fontenc}

\usepackage{color,soul}

\usepackage[dvipsnames]{xcolor}
\definecolor{light-gray}{gray}{0.95}

\usepackage{csquotes}
\usepackage{microtype}

\begin{document}

\title{The LLM Mirage: Economic Interests and the Subversion of Weaponization Controls}

\author{Ritwik Gupta}
\email{ritwikgupta@berkeley.edu}
\orcid{0000-0001-7608-3832}
\author{Andrew W. Reddie}
\email{areddie@berkeley.edu}
\orcid{0000-0003-3231-8307}
\affiliation{%
  \institution{University of California, Berkeley}
  \city{Berkeley}
  \state{CA}
  \country{USA}
}

\renewcommand{\shortauthors}{Gupta and Reddie}

\begin{abstract}
U.S. AI security policy is increasingly shaped by an \emph{LLM Mirage}, the belief that national security risks scale in proportion to the compute used to train frontier language models. That premise fails in two ways. It miscalibrates strategy because adversaries can obtain weaponizable capabilities with task-specific systems that use specialized data, algorithmic efficiency, and widely available hardware, while compute controls harden only a high-end perimeter. It also destabilizes regulation because, absent a settled definition of ``AI weaponization,'' compute thresholds are easily renegotiated as domestic priorities shift, turning security policy into a proxy contest over industrial competitiveness. We analyze how the LLM Mirage took hold, propose an intent-and-capability definition of AI weaponization grounded in effects and international humanitarian law, and outline measurement infrastructure based on live benchmarks across the full AI Triad (data, algorithms, compute) for weaponization-relevant capabilities.
\end{abstract}




\maketitle

\section{Introduction}
On January 15, 2025, the U.S. Department of Commerce issued the ``Framework for Artificial Intelligence Diffusion,''~\cite{bureauofindustryandsecurityFrameworkArtificialIntelligence2025} extending hardware export controls into a broader regime that treated training compute and chip performance as the administrative boundary of ``frontier'' AI. Within four months, and in the context of a change of administration, Commerce announced it would rescind the framework, describing it as burdensome for industry and damaging to diplomatic relations~\cite{bureauofindustryandsecurityDepartmentCommerceAnnounces2025}. Later, the administration approved high-end accelerator exports to select customers in China under a revenue-sharing arrangement~\cite{nellisUSAllowNvidia2025, hawkinsUSAllowPowerful2025}.

While this sequence may seem like policy volatility across administrations at a surface level, it instead reveals a fundamental measurement problem. U.S. AI security governance has not settled on a durable definition of the core societal harm it is trying to prevent. In the gap between executive-branch warnings about chemical, biological, radiological, and nuclear (CBRN); cyber misuse; and the Commerce Department's compute thresholds, ``weaponization'' is often implied rather than specified. As the harm target was underspecified, the proxy of compute steadily became the objective.

We call this dynamic the \textbf{LLM Mirage}: the assumption that the danger an AI system poses is determined primarily by how much computing power was used to build it, and that restricting access to that computing power is therefore sufficient to prevent misuse.

The costs show up in two places. Strategically, compute-centric controls harden a boundary around the most resource-intensive models while leaving plausible weaponization vectors---task-specific targeting, autonomy, and cyber operations---available on less restricted hardware through optimized algorithms and curated data. Politically, once the regulated object is ``frontier infrastructure,'' governance concentrates agenda-setting power in the hands of infrastructure incumbents.

This work contributes to the FAccT community's ongoing investigation into the politics of AI quantification and governance. By analyzing how compute thresholds function as a classification regime, we reveal how national security is used to justify the consolidation of infrastructure power. We sit at the intersection of accountability and transparency, arguing that current policy ignores the distributional harms of silicon-centric security.

Concretely, in this paper we
\begin{enumerate}
    \item diagnose compute-centric AI security policy as a proxy-driven governance failure (\Cref{sec:background});
    \item propose an effects-based definition of AI weaponization grounded in intent, demonstrated capability, and International Humanitarian Law (\Cref{sec:definition}); and
    \item operationalize that definition through live, adversarial benchmarks that evaluate the full AI Triad and support defensible updates to export-control and disclosure thresholds (\Cref{sec:operationalizing}).
\end{enumerate}

\section{Background}
\label{sec:background}

Compute-centric governance persists because it fits previously reliable export control enforcement mechanisms. Agencies write export controls as thresholds, enforce them through shipment licensing, and audit them through infrastructure. That fit, however, carries an analytic cost. It turns the contested question of what kinds of AI use constitute weaponization risk into an infrastructure question about what can be counted and controlled. This section traces how compute became a plausible shorthand for ``frontier'' capability, and why governing through that shorthand predictably produces both strategic blind spots and political instability.

\subsection{How scaling made compute look decisive}

The performance of AI models reflects the interaction of \textit{data}, \textit{algorithms}, and \textit{compute}. The research community has long debated which input matters most in practice. However, a sequence of empirical regularities made compute scaling feel unusually reliable for producing broad, general-purpose performance. That perception matters because it later became the intellectual justification for treating compute as both a research input and a national security lever.

Early work emphasized data. Halevy, Norvig, and Pereira~\cite{halevyUnreasonableEffectivenessData2009} highlighted the ``unreasonable effectiveness'' of data. Sufficiently large datasets can allow comparatively simple methods to perform well. They asserted that while hand-crafted improvements to algorithms matter, more predictable gains can be acquired by scaling the amount of data simple algorithms operate on top of. For governance, this reinforced the idea that ``frontier progress'' could be summarized by a single dominant input when measurement and enforcement demanded simplification.

Further, Sutton's ``Bitter Lesson''~\cite{suttonBitterLesson2019} observed that over long horizons, general learning methods that leverage search absorb more computation and more experience, which then outperform approaches that rely on specialized human insight. In practice, this argument reinforced a research strategy that prioritized scalable training pipelines and general architectures. It also nudged the field toward a view in which algorithmic novelty matters, but scaling is the metronome of progress. In the dominant ``frontier'' governance conversations, \textit{The Bitter Lesson} shifted the default story from algorithmic ingenuity to computational scale.

Deep learning researchers formalized these intuitions through research establishing empirical scaling relationships. Researchers at Baidu articulated power-law behavior showing that performance can scale predictably with increases in model size and training data~\cite{hestnessDeepLearningScaling2017}. Work from OpenAI extended and popularized this idea for Transformer-based models, establishing correlations among model size, dataset size, and training compute (measured in petaFLOPs)~\cite{kaplanScalingLawsNeural2020}. In sum, these works showed that systematic gains could be achieved by increasing scale across multiple dimensions in tandem---and that compute budgets offered the most predictive measure of that scale.

Later results refined the perception. Work on the Chinchilla family of models argued that given a fixed compute budget, there is a compute-optimal tradeoff between parameters and training tokens~\cite{hoffmannTrainingComputeOptimalLarge2022}. This strengthened a practical belief that compute allocations govern the feasible frontier. If one can secure large compute budgets and allocate them efficiently, one can reach higher performance with some predictability. In parallel, the assertion of ``emergent abilities''---capabilities that appear in large models but are absent in smaller ones and are not straightforwardly predictable from extrapolation---added urgency to the notion of threshold effects~\cite{weiEmergentAbilitiesLarge2022}. If some capabilities arrive discontinuously, then it becomes politically and strategically natural to fear ``crossing the line,'' and to treat reaching (or denying) the compute needed to cross that line as central. In aggregate, the discourse around scaling laws turned compute budgets into an unusually legible single-number summary of ``frontier-ness,'' despite a deeply complex underlying causal.

At the same time, researchers contested the scaling paradigm even as it hardened into orthodoxy. Prior work has warned that equating parameter count or training compute with reasoning capability is a category error~\cite{hookerLimitationsComputeThresholds2024, guptaDataCentricAIGovernance2024}. Related critiques emphasized the lack of semantic understanding in purely statistical language models, the measurement validity problems associated with the correlation of benchmark performance with competence, and even the environmental costs and externalities of large-scale training~\cite{benderDangersStochasticParrots2021, narayananAISnakeOil2024}. These critiques emphasize that even if scale is correlated with some forms of performance in frontier LLM development, scale itself is not a stable substitute for understanding what a system can actually do in context, or what harms it can enable.

Against this technical backdrop, national security actors began translating ML intuitions into policy. Ben Buchanan labeled data, algorithms, and compute as the ``AI Triad'' and argued that compute offered the United States a uniquely durable asymmetric advantage because advanced semiconductors are concentrated and exportable~\cite{buchananAITriadWhat2020}. In his framework, data was comparatively ``overvalued and overhyped'' as a strategic choke point, while compute was portrayed as enforceable leverage. This interpretation proved influential because it aligned with both the scaling narrative (compute as frontier shorthand) and the institutional realities of enforcement (chips as inspectable, licensable objects). But it also set up the central substitution this paper critiques: the move from ``compute is a practical bottleneck for frontier research'' to ``compute is a proxy for weaponization risk.''

\subsection{How scaling became a regulatory proxy}

Compute's legibility as a shorthand made it easy to translate into policy. Intent, deployment context, and downstream weaponization vectors are hard to measure and even harder to govern. Compute is measurable. It can be linked to physical supply chains, licensing regimes, and reporting requirements. As a result, U.S. AI security policy increasingly defined the ``frontier'' through input thresholds---training FLOPs, cluster size, and chip performance---instead of observed effects in real-world operational settings.

This translation is visible in President Biden's Executive Order 14110~\cite{whitehouseSafeSecureTrustworthy2023}. The order institutionalized a focus on scaling by creating a regulatory framework centered on ``dual-use foundation models,'' defined as AI models that pose serious national security risks. Critically, it defined the frontier through high-water marks tied to model size and compute, rather than a task-based assessment. The order established a bright-line threshold for models trained with more than $10^{26}$ integer or floating-point operations and noted that covered models contain ``at least tens of billions of parameters.'' The order justified this focus by pointing to CBRN and cyber risks, including lowering barriers for non-experts to design or use CBRN weapons and enabling powerful offensive cyber operations. In other words, the governance logic was effects-motivated, but the regulatory trigger was input-oriented.

The Department of Commerce's export controls further reinforced the same logic. The October 2022 controls restricted high-end chips and related technologies~\cite{bureauofindustryandsecurityImplementationAdditionalExport2022}. These controls aimed to slow China's military modernization by limiting access to the hardware underpinning advanced AI. Again, the frontier of AI hardware was defined through quantitative performance thresholds, specifically total processing power and high-speed interconnect bandwidth, that can be enforced through licensing and shipment controls. This effectively treated cutting-edge compute as the primary strategic chokepoint, privileging control of the newest and most powerful chips as a pathway to controlling advanced AI systems.

The January 2025 ``Framework for Artificial Intelligence Diffusion'' extended the compute-centric approach beyond hardware into the governance of model distribution~\cite{bureauofindustryandsecurityFrameworkArtificialIntelligence2025}. The framework sought to manage the global spread of advanced AI by regulating both high-end chips and the export of trained model weights. It proposed a tiered system in which trusted allies received streamlined access, nations of concern faced strict denials, and other countries were assigned compute-based quotas. Yet, despite this geographically differentiated architecture, the framework retained compute as its definitional core. The amount of training compute primarily determined which models triggered restrictions rather than by demonstrated use conditions and effects.



\subsection{Measurement politics and capture}
Compute thresholds do more than provide a proxy measurement for risk; they operationalize a theory of what risk is. Quantification makes some phenomena legible to governance while pushing others to the margins~\cite{espelandSociologyQuantification2008, merrySeductionsQuantificationMeasuring2016}. In the context of AI security, defining the ``frontier'' through training FLOPs, chip performance, or cluster scale privileges what can be inspected at ports and audited in data centers over intent, operational integration, doctrine, and effects in context.

Thresholds also function as classification systems. Once a rule sorts systems into regulated and unregulated categories, those categories take on institutional reality even when the underlying harms are heterogeneous and contingent~\cite{bowkerSortingThingsOut2000}. The proxy becomes especially brittle when it is treated as the object of governance rather than as a partial indicator of a broader construct.

This is a familiar failure mode in metric-centered regulation. When a measure becomes a target, actors game it in ways that weaken its relationship to the construct it was meant to represent. Industry-funded safety benchmarks can become a form of "transparency theater" \cite{douekContentModerationSystems2022, rajiOutsiderOversightDesigning2022}, where compute-centric audits allow incumbents to self-grade while excluding the public from defining what counts as a risk. In AI security policy, that gap has significant effects. Weaponization depends on capability, intent, and deployment context, not on a single scalar input. Treating compute as a one-number proxy invites both false negatives (harmful systems below the threshold) and false positives (high-compute systems outside weaponization pathways).

Finally, measurement choices reallocate influence. When compliance and evaluation hinge on assets controlled by a small set of firms, those firms gain structural agenda-setting power over what evidence is produced, what risks are treated as salient, and what oversight is considered feasible~\cite{weiHowAICompanies2025}. Where evaluators lack meaningful access and must rely on firm-mediated interfaces, the most ``auditable'' actors can shape what gets audited and how~\cite{terzisLawEmergingPolitical2024, youngConfrontingPowerCorporate2022}.

\section{Objective Drift}
\label{sec:drift}

The consequences of compute-centric governance manifest along two distinct but reinforcing dimensions. First, in the absence of an operational definition of what AI controls are intended to prevent, policy objectives become unstable and susceptible to drift as debates shift from threat mitigation to industrial and diplomatic bargaining. Second, even when controls remain formally intact, a perimeter defined by compute thresholds provides misplaced reassurance by leaving open the pathways through which weaponizable capabilities actually emerge. Below, we examine these dynamics in turn.

\subsection{From mitigating threats to industrial competitiveness}
Compute-linked controls translated diffuse national security concerns into an administrable rule. That translation, however, still leaves Washington without a testable target condition for what the rules are meant to stop. In that gap, justification can drift from threat mitigation toward industrial competitiveness, ultimately resulting in negotiation over infrastructure.

The rescission of the diffusion framework illustrates this dynamic. Commerce emphasized burdens on U.S. industry and effects on diplomacy rather than engaging the concrete weaponization risks and use cases the rule was meant to constrain~\cite{bureauofindustryandsecurityDepartmentCommerceAnnounces2025}. Infrastructure-focused governance, which is inherently tied to large capital investments, makes it easy for invested parties to pivot the debate towards deals and industrial advantage over evidence of capability and use.

Indeed, the rollback did not stop at the reversal of the AI Diffusion Framework. In December 2025, the Trump administration authorized exports of NVIDIA's high-end H200-class AI accelerators to approved customers in China under a revenue-sharing arrangement~\cite{nellisUSAllowNvidia2025, hawkinsUSAllowPowerful2025}. Because the policy community never operationally specified ``weaponization,'' no stable evidentiary burden constrained its reversal. Instead, the debate became a negotiation over private sector interests and competitiveness rather than a test of credible misuse scenarios, and ``security'' became a label attached to whatever threshold was convenient.

\subsection{Bypassing a compute-defined perimeter}
Beyond instability, the more significant strategic cost of the LLM Mirage is misplaced reassurance. A compute-defined perimeter can look stringent while leaving open the pathways that matter for operational weaponization. Two patterns are already visible: (1) capability advances through algorithmic efficiency on export-compliant hardware; and (2) weaponizable task-specific systems that sit well below ``frontier'' thresholds.

First, Chinese firms have trained state-of-the-art models such as \textit{Hunyuan}-Large~\cite{sunHunyuanLargeOpenSourceMoE2024} and DeepSeek-R1~\cite{deepseek-aiDeepSeekR1IncentivizingReasoning2025} on export-control-compliant GPUs by leveraging common modeling and training techniques~\cite{guptaWhackaChipFutilityHardwareCentric2024}. In that setting, static hardware thresholds constrain headline infrastructure without reliably bounding the development of weapons.

Second, specialized systems can deliver militarily salient effects without frontier compute. The Ukrainian military is deploying small models with millions of parameters ($\sim1000\times$ smaller than the smallest LLMs) on modest hardware (NVIDIA Jetson Orins, roughly 14$\times$ less powerful than an NVIDIA H200) at scale~\cite{bondarUkrainesFutureVision2025}. On the other side of the world, the Chinese-origin \textit{Zhousidun} dataset was created to train an object detection model designed to identify and target radar and missile components on U.S. and allied naval destroyers~\cite{guptaOpenSourceAssessmentsAI2024}. These cases are exemplars of pathways a compute-centric regime tends to underweight.

Together, these patterns point to a measurement problem more so than an enforcement problem. Governing weaponization requires evaluating demonstrated performance and use conditions, not assuming that compute levels delimit operational risk.

\section{Distributional Consequences}
\label{sec:distributional}
Compute-centric security governance mismeasures risk and redistributes access, influence, and exposure. The LLM mirage functions as a classification regime that allocates privileges and burdens through a proxy that correlates with capital intensity. The result is a security policy that can appear neutral while producing patterned winners and losers.

\subsection{Access is rationed by capital instead of demonstrated risk}
Tiered export regimes and compute thresholds sort countries, firms, and research institutions by their position in supply chains and alliance networks rather than by demonstrated weaponization pathways. The diffusion framework's tier architecture made this explicit by creating differentiated access to advanced chips and model assets across geopolitical categories, with many destinations governed through quantitative caps rather than capability-specific evidence. Within the United States and allied ecosystems, compute-triggered compliance costs also scale with organizational capacity. Large firms can absorb reporting, secure facilities, and evaluator engagement; smaller labs and public-interest researchers often cannot.

\subsection{Risk exposure shifts toward Global South populations}
The LLM Mirage enforces a specific geopolitical epistemology of security. In international relations, ``securitization'' occurs when an issue is framed as an existential threat to justify extraordinary measures---such as the compute-centric export controls and diffusion frameworks analyzed in this paper \cite{buzanSecurityNewFramework1998}. In the context of AI, this securitization is predominantly led by Global North institutions, which define ``safety'' and ``security'' through the lens of frontier parity and existential risk.

Current governance regimes rely on a security-through-scarcity model. However, for nations in the Global South, this safety manifests as enforced technological underdevelopment \cite{safirDistributiveEpistemicInjustice2025}. When compute thresholds are set as the primary trigger for regulatory oversight, they become a barrier to entry for local innovation, effectively forcing dependency on Global North software-as-a-service or infrastructure-as-a-service offerings.

Global South populations face immediate, task-specific weaponization and surveillance risks which the LLM Mirage overlooks. As non-LLM-based surveillance tools are exported to African, Southeast Asian, and Latin American states~\cite{jiliDigitalSovereigntyPostcolonial2025} under the guise of technological partnership, they create new paradigms of strategic vulnerability and eroded algorithmic sovereignty. In these contexts, security is a lived reality of automated over-policing, biometric exclusion, and the suppression of political dissent facilitated by imported AI systems that operate without local accountability or contextual relevance \cite{heeksDigitalDivideDigital2020}.

\section{Defining AI Weaponization}
\label{sec:definition}

To construct a regulatory framework for artificial intelligence that is both durable and effective, we must first fill the vacuum that the LLM Mirage currently occupies. The prevailing policy discourse operates on a definition of weaponry rooted in the industrial age, fixated on physical inputs and computational scale. To correct this, we propose a scoped, effects-based definition of AI weaponization. In the sections below, we first articulate this definition, and then utilize it to align the legal frameworks of International Humanitarian Law (IHL) and the strategic realities of 21st-century conflict.

\subsection{An intent-and-capability definition of AI weaponization}
We propose the following definition:

\begin{displayquote}
\emph{
    AI weaponization is the \textcolor{BurntOrange}{intentional employment or objective immediate capability} of an artificial intelligence system to cause \textcolor{LimeGreen}{physical, digital, or psychological harm}, to create an \textcolor{RoyalBlue}{asymmetric advantage} in conflict, or to \textcolor{Thistle}{materially facilitate} such effects.
}
\end{displayquote}

\subsubsection{\textcolor{BurntOrange}{Intentional employment or objective immediate capability}}
This anchors the definition in the realities of modern proliferation. The first component targets the actor by capturing the deliberate misuse of AI (\textit{mens rea}), distinguishing between a terrorist group's misuse and a researcher's legitimate study of toxicology. However, intent is difficult to prove and insufficient for prevention. Therefore, the second component establishes that a system capable of designing novel pathogens or generating zero-day exploits constitutes a weaponized capability by virtue of its performance, regardless of the developer's stated intent.

\subsubsection{\textcolor{LimeGreen}{Physical, digital, or psychological harm}}
This extends the definition beyond kinetic effects to encompass the primary domains of 21st-century conflict. ``Physical harm'' refers to traditional kinetic violence and destruction. ``Digital harm'' addresses offensive cyber operations, such as data destruction, critical infrastructure disabling, or network disruption. ``Psychological harm'' acknowledges the weaponization of the information domain, specifically systems capable of generating targeted propaganda, deepfakes, or disinformation at a scale designed to incite physical violence, disrupt critical civic processes, or simulate official communications.

\subsubsection{\textcolor{RoyalBlue}{Asymmetric advantage}}
Asymmetric advantage allows a smaller force to effectively neutralize a larger adversary~\cite{mackWhyBigNations1975, paulAsymmetricConflictsWar1994}. This element narrows the scope to capabilities that plausibly change wartime outcomes, rather than routine commercial efficiencies. This makes the definition applicable to AI systems used for intelligence analysis, logistics optimization, or electronic warfare. A model that shifts the offense-defense balance, even through non-lethal means, constitutes a weaponized application. This element ensures that regulation remains focused on capabilities that pose a genuine threat to national security and stability, distinguishing them from standard commercial efficiencies.

\subsubsection{\textcolor{Thistle}{Materially facilitate}}
This clause extends the definition to systems that materially reduce the skill, time, or resources required to carry out harm. An AI system need not pull a trigger to be weaponized; it creates equal strategic risk if it compresses the time, skill, or resources required for an adversary to launch an attack. Examples include models that design viable synthesis pathways for nerve agents or autonomously discover software vulnerabilities. To distinguish weaponization from benign research, the downstream impact must be immediate or achievable with low-resource modifications. A specialized system, such as a crop-monitoring model that requires substantial retraining or architectural modification to identify military targets, does not meet this threshold, even if it holds latent dual-use potential. This boundary is meant to preserve room for legitimate dual-use research while still flagging systems whose marginal modifications predictably enable high-consequence misuse.

\vspace{\baselineskip}

This definition is intentionally independent of computational scale. It can treat a small, specialized system trained on curated data as weaponizable when it demonstrates effectiveness on high-consequence tasks, and it can exclude a large foundation model used for benign purposes when credible misuse scenarios are not present.

While our definition of weaponization utilizes the language of IHL, we acknowledge that security itself is a contested political construct that disproportionately harms the Global South. Through our proposed definition, by focusing on intended effects rather than computational inputs, we can move away from the security-through-scarcity model that systematically removes the Global South from international AI governance.



\subsection{From physical instrument to intended effect via International Humanitarian Law}
This intent-and-capability standard proposed above is preferable because it abandons domestic statutes' input-based regulation in favor of the epistemological framework of International Humanitarian Law (IHL)~\cite{diplomaticconferenceonthereaffirmationanddevelopmentofinternationalhumanitarianlawapplicableinarmedconflictsProtocolAdditionalGeneva1977}. While domestic laws and export control regimes frequently define weapons by their physical mechanisms---such as firearms that ``expel a projectile by the action of an explosive''---the law of armed conflict offers a more sophisticated, effects-based standard. IHL recognizes that the legal and strategic significance of a weapon lies not in its material form, but in the effects it produces and the manner in which it is used. Unlike a tank or a missile, an AI system's weaponization potential is rooted more in its operational integration than its hardware footprint.

Under IHL, states regulate weapons not simply as objects, but as means and methods of warfare. This distinction, embedded in Additional Protocol I to the Geneva Conventions, directs legal scrutiny toward whether a weapon or method of warfare is capable of being employed in compliance with core principles such as distinction, proportionality, and military necessity. Article 36 weapons reviews, in particular, require states to assess whether the use of a new weapon, means, or method of warfare would be prohibited under international law. Such assessments eschew physical composition alone and instead additionally encompass effects, foreseeable use, and operational context.

This framing accommodates non-traditional weapons. States regulate chemical and biological agents, cyber capabilities, and certain forms of information operations not because of their physical characteristics, but because of their effects on combatants, civilians, and the conduct of hostilities~\cite{priceGenealogyChemicalWeapons1995, tannenwaldNuclearTabooUnited1999, bentleyBiologicalWeaponsTaboo2024, muellerTotemTabooDepolarizing2003}. A virus can be a weapon; code can be a weapon; a commercial system can become a weapon through either an intended or unintended use. What matters under IHL is not whether harm is delivered kinetically, but whether a tool is intentionally employed to cause damage or gain military advantage in a manner that implicates humanitarian protections.

Weapons law scholars have developed extensive frameworks for applying IHL's principles of distinction and proportionality to novel means of warfare, such as autonomous systems~\cite{schmittOutLoopAutonomous2013}, demonstrating that effects-based legal review is both required and practically implementable even when a system lacks conventional physical form~\cite{cornAutonomousWeaponSystems2014}.

Artificial intelligence systems fit squarely within this effects-based legal tradition. An AI model is not a weapon by virtue of its architecture or training compute, but it can become one through the intentional development, modification, or deployment of the system to produce military effects. An AI system used for automatic target recognition, autonomous navigation of munitions, cyber exploitation, or large-scale psychological operations functions as a means of warfare regardless of whether it is embedded in a missile, a drone, or a data center~\cite{burtonUnderstandingStrategicImplications2019, linArtificialIntelligenceNuclear2025, schwartzOutLoopAgain2025, kahnRiskyIncrementalismDefense2024}. Treating such systems as outside the scope of weapons regulation because they lack a conventional physical form misreads both the letter and the spirit of IHL.

This emphasis on effects instead of platforms is mirrored in contemporary military doctrine. U.S. joint doctrine---and Air Force doctrine in particular---explicitly frames warfare as ``effects-based,'' prioritizing outcomes over delivery mechanisms~\cite{deptulaEffectsBasedOperationsChange2001}. Whether a target is neutralized by a manned aircraft, an uncrewed system, or a cyber operation is operationally secondary to the effect achieved. Modern joint warfighting is organized around integrating capabilities across domains to produce desired strategic results, not around the intrinsic properties of individual platforms.

Current compute-centric AI governance sits uneasily with IHL's effects-based logic and risks leaving genuinely weaponized AI systems outside meaningful regulatory scrutiny while over-regulating benign technologies. Indeed, pushing the frontier should be encouraged rather than discouraged from a technology competitiveness perspective. A durable framework for AI governance should therefore align with the established principles of the law of armed conflict and regulate AI systems based on their intended use and foreseeable effects, not the hardware on which they are trained.

\subsection{The strategic logic of conflict}
The necessity of an effects-based definition of weaponization becomes clear when viewed through the lens of cost-of-force exchange ratios~\cite{fowlerCostExchangeRatio1997}. In contemporary great-power competition, strategic advantage increasingly accrues not to the actor fielding the most exquisite platforms, but to the actor that can impose unfavorable exchange ratios on its adversary over time. Artificial intelligence represents one technology (in tandem with others) that is rapidly changing this dynamic by enabling relatively inexpensive systems to identify, target, and destroy far more costly ones~\cite{mazarrDefendingDominanceAccelerating2023}.

Modern conflict increasingly revolves around attrition under conditions of technological asymmetry, even among advanced militaries. Today's militaries are increasingly seeking to leverage low-cost, AI-enabled unmanned systems, often assembled from commercial components and guided by relatively simple models~\cite{scharreArmyNoneAutonomous2019}, to destroy far more expensive systems, whether tanks, air defense systems, or aircraft carriers costing orders of magnitude more. In many cases, the strategic effect is not the individual tactical loss, but the cumulative erosion of force structure and the unsustainable economics imposed on the defender~\cite{horowitzWhenSpeedKills2019}. When a system costing thousands of dollars can reliably destroy one costing millions, traditional assumptions about deterrence, escalation control, and military modernization break down.

For the United States, whose force structure is heavily weighted toward high-end, capital-intensive platforms---aircraft carriers, fifth-generation fighters, advanced missile defense systems, and its nuclear arsenal---this dynamic is particularly dangerous~\cite{smithDefenceAcquisitionProcurement2022}. The strategic risk is not that adversaries will outcompete the United States in building more exquisite systems, but that they will instead exploit AI-enabled sensing, targeting, and coordination to make U.S. platforms increasingly vulnerable to cheaper means of attack~\cite{eslamiGeopoliticsAIDrivenArms2025}. In this context, the most consequential AI applications are not frontier large language models, but task-specific systems that compress the operational timeline: automatic target recognition, sensor fusion, trajectory optimization, and real-time battle damage assessment.

This logic fundamentally challenges compute-centric AI governance~\cite{robinsonEstablishmentInternationalAI2025, hatzNuclearAnalogyAI2025, kopAtomicAgencyQuantumAI2025, chestermanTragedyAIGovernance2024, emery-xuInternationalGovernanceAdvancing2025}. Weaponization, particularly in a great-power context, is not defined by the sophistication or scale of an AI system in isolation, but by its ability to alter force-exchange dynamics. A small, specialized model that improves the probability of kill for a loitering munition, or that enables cooperative targeting among swarms of inexpensive drones, can have greater strategic impact than a vastly more powerful general-purpose model used for benign applications. What matters is whether AI lowers the marginal cost of imposing damage on an adversary's most valuable assets.

Importantly, this dynamic does not require technological parity. Adversaries need not match U.S. investment in compute, data centers, or frontier models to achieve destabilizing effects. They need only achieve \textit{sufficient} capability to exploit vulnerabilities in U.S. force design and acquisition assumptions. AI accelerates learning, iteration, and adaptation, allowing militaries to rapidly refine tactics and systems that optimize for unfavorable exchange ratios. This is precisely why focusing regulatory attention on the apex of the AI ecosystem obscures the systems most likely to be weaponized in practice.

Viewed through this lens, the defining characteristic of system as a weapon depends on its ability to shift the economics of conflict, a functional reality that remains independent of its underlying computational pedigree. An AI system is strategically significant if it allows one actor to destroy, disable, or neutralize an adversary's forces at a fraction of the cost the adversary must pay to defend or replace them. 

A regulatory framework that ignores this reality---by equating risk with model scale or training compute---will fail to address the most probable pathways through which AI reshapes great-power military competition. From our perspective, this is a significant blindspot in the contemporary policy discussions.

\section{Operationalizing Definitional Clarity into Policy}
\label{sec:operationalizing}

An effects-based definition of AI weaponization is only useful if it can be put into practice. Doing so requires shifting away from input proxies and towards empirical evaluation of what AI systems can actually do under realistic conditions. As such, the core task is to identify and measure the specific capabilities that would constitute weaponization in practice, rather than assuming that risk scales linearly with compute, model size, or training cost. We propose an iterative framework in which weaponization-task-specific benchmarks are used to evaluate each axis of the AI triad, the results of which are used to set dynamic thresholds for controls on software, hardware, or a combination of both.

A benchmark-driven regime also raises the question of who decides what counts as dangerous capability, and who can contest those decisions? To reduce capture risk, evaluation infrastructure should be independent of the firms being assessed, transparent about benchmark construction and aggregate results (with narrowly scoped security redactions), and governed through mechanisms that include independent researchers and civil society alongside government and industry. These safeguards do not eliminate politics, but they make it harder for frontier incumbents to define ``security'' as whatever preserves their advantage.

\subsection{Comprehensive weaponization benchmarks}

The first step is to develop a granular taxonomy of high-risk capabilities grounded in concrete national security concerns. While not exhaustive, three categories illustrate the most salient risks motivating contemporary AI policy: CBRN proliferation, offensive cyber operations, and autonomous or semi-autonomous military functions (particularly where there are kinetic effects).

In the CBRN domain, the primary concern is not passive information access but the use of generative and optimization-based models to design novel toxic chemicals or biological agents with tailored properties~\cite{barrettBenchmarkEarlyRed2024, kumarQuantifyingCBRNRisk2025}. Prior research~\cite{urbinaDualUseArtificialintelligencepowered2022} has demonstrated that AI systems developed for drug discovery can be readily repurposed to generate thousands of potentially lethal compounds, including many that fall outside existing regulatory watchlists. The risk lies not in factual recall, but in procedural enablement, i.e., compressing expertise, accelerating iteration, and lowering barriers to synthesis.

In offensive cyber operations, high-risk capabilities include the use of AI agents to identify vulnerabilities in large codebases, discover zero-day exploits, or autonomously generate functional exploit code~\cite{sinhaFirewallsFrontiersAI2025}. These tasks directly reduce the time, skill, and organizational capacity required to conduct sophisticated cyber attacks~\cite{anthropicDisruptingFirstReported2025}, with clear implications for critical infrastructure and military networks.

Military-relevant weaponization includes capabilities such as automatic target recognition in cluttered or degraded environments, autonomous navigation and coordination under adversarial conditions, and large-scale generation of tailored propaganda or deepfakes intended to erode public trust or manipulate decision-making. These functions are destabilizing in part because they operate in gray-zone contexts below traditional thresholds of armed conflict.

Identifying such capabilities is a necessary first step which further requires measurement. This, in turn, demands a suite of concrete, task-specific benchmarks. In this framework, a benchmark is a controlled evaluation defined by four components: the datasets used, the performance metrics applied, the operational context of the test, and the hardware on which the system is evaluated. Since benchmarks can become \textit{de facto} policy instruments, they should be treated as measurement claims rather than as neutral tests. Each benchmark should make explicit what construct it is intended to measure (and what it does not), what forms of external validity are plausible, and how results should or should not generalize across deployment conditions.

For example, a CBRN benchmark would not ask whether a model can describe a nerve agent, but whether it can design a novel toxic molecule or propose a viable synthesis pathway. Evaluation would draw on established chemical and biological databases (e.g., ZINC, PubChem) and assess outputs using effects-based metrics such as predicted toxicity, novelty, and synthesis feasibility~\cite{moutonOperationalRisksAI2024}. Contextual variation---such as access to external simulation tools or deployment on commodity hardware---would test whether dangerous capabilities emerge outside frontier compute environments.

Similarly, cyber benchmarks would evaluate a model's ability to generate working exploits in sandboxed environments using vulnerable codebases like the Juliet Test Suite~\cite{richardsonArichardsonJuliettestsuitec2025}. Performance would be measured by time-to-discovery and exploit success rates under realistic access constraints (e.g., black-box versus white-box conditions) rather than descriptive accuracy.

In the military domain, benchmarks could assess automatic target recognition or autonomous coordination using imagery datasets such as xView~\cite{lamXViewObjectsContext2018} or high-fidelity simulators. Metrics would be tied to mission-relevant outcomes---e.g., precision and recall under sensor degradation, robustness to electronic warfare conditions, or task completion despite simulated losses---instead of laboratory accuracy alone.

By evaluating how specific combinations of data, algorithms, and compute translate into real-world effects, this benchmark-driven approach replaces static proxies like FLOPs with direct evidence of risk. Further, it allows regulators to identify weaponizable capabilities regardless of whether they arise from frontier-scale models or small, specialized systems trained on curated data.

Early efforts in this direction already exist. The Weapons of Mass Destruction Proxy (WMDP) benchmark, for instance, assesses whether models contain hazardous domain knowledge across biosecurity, chemical security, and cybersecurity~\cite{liWMDPBenchmarkMeasuring2024}. However, WMDP primarily measures knowledge acquisition, not procedural competence. A model can correctly answer detailed questions about chemical precursors without being able to design a viable synthesis route or delivery mechanism. For regulatory purposes, this distinction is decisive. Governance frameworks must therefore move beyond asking what models know to rigorously testing what they can do. Conversely, others have built crisis simulations that measure the adherence of LLMs towards IHL principles, but stop short of building a reproducible evaluation~\cite{drinkallRedLinesGrey2025}.

\subsection{Integrated, live benchmarks for the AI triad}

Any actor seeking to deploy or export advanced capabilities along any axis of the AI triad should be required to demonstrate that those capabilities do not materially increase risk. We propose a regulatory framework that evaluates how specific combinations of resources translate into concrete, weaponizable capabilities.

For hardware manufacturers, export licensing for new compute platforms would be contingent on providing limited access to the relevant hardware for controlled evaluation. This enables direct testing of whether novel compute architectures meaningfully accelerate high-risk capabilities relative to existing systems. Similarly, firms that control large, proprietary datasets intended for generative AI training would be required to make a representative subset available for secure benchmarking, allowing evaluators to assess whether data access alone materially shifts the risk frontier.

The framework would then incentivize academic researchers and trusted third parties to develop state-of-the-art algorithms using these shared resources. Resulting systems would be continuously evaluated against a standing suite of high-risk task benchmarks. This live evaluation process serves two functions. First, it provides an empirical measure of the pace of model development, allowing regulators to maintain a calibrated understanding of the adaptation buffer. Second, it generates quantitative evidence to justify regulatory intervention. Policy can respond as soon as benchmarks show that the Pareto frontier of dangerous capabilities has advanced---regardless of whether that advance is driven by compute, data, or algorithms.

\subsection{A New Institutional architecture}
We propose a hybrid institutional architecture that leverages existing U.S. government authorities to realize our live benchmarking framework. The National AI Research Resource (NAIRR) should serve as the infrastructure backbone, providing a secure, air-gapped computational enclave (NAIRR Secure) where proprietary models can be hosted for evaluation without risking IP leakage. However, the governance and testing mandate must reside with the Center for AI Standards and Innovation (CAISI) at NIST. As the agency responsible for standards and measurement, CAISI possesses the requisite technical authority to design validity-tested benchmarks for CBRN and cyber capabilities. Under this division of labor, NAIRR provides the public-interest audit infrastructure, while NIST provides the testing harness.

Positioning NAIRR Secure as the locus of live evaluation offers several advantages. It centralizes sensitive testing within a trusted, publicly accountable institution and does not place the sole burden on private firms with conflicting incentives. It ensures that the United States' understanding of the AI frontier is not exclusively shaped by corporate disclosures or adversary behavior. It provides a consistent pipeline of up-to-date hardware for researchers building on NAIRR. Lastly, it creates a shared empirical baseline from which export controls, licensing decisions, and release thresholds can be justified and defended.

If live benchmarking reveals that a dangerous application (e.g., automated vulnerability discovery) can be achieved on a cluster of mid-tier GPUs using optimized algorithms, the CAISI/NAIRR architecture must trigger an immediate update to Commerce Department thresholds. In this model, benchmarks serve as upstream intelligence for downstream enforcement: they determine the technical specifications of the thresholds based on the reality of specific threats, ensuring that export controls restrict the hardware actually required for weaponization, not just the hardware required for commercial prestige.

\section{Broader Impacts and Limitations}
This paper analyzes U.S.-led compute-centric governance as a case of measurement-driven regulation under geopolitical pressure. We focus on national security claims about weaponization (CBRN, cyber, military targeting) rather than broader civilian harms. We do not claim that compute is irrelevant; we argue that compute thresholds are an unstable substitute for defining and testing weaponization.

Benchmark-driven governance can reduce instability, but it introduces its own risks. Centralized evaluation can become a bottleneck that advantages actors with privileged access to testing capacity. Benchmarks can be optimized against or shaped by contested value choices about what is measured and what is ignored. A benchmark regime can further legitimate a narrow, state-centric account of ``security,'' crowding out civilian harms and obscuring which communities bear downstream exposure.

These risks strengthen the case for narrow mandates, transparent benchmark construction and validity claims, meaningful contestation procedures, and institutional separation between evaluators and commercial AI system developers.

\section{Escaping the LLM Mirage}
U.S. AI governance has treated computational scale as a stand-in for weaponization risk. That choice is administratively convenient, but it weakly targets the pathways that drive operational misuse, and it leaves policy politically fragile once security controls are interpreted as tools of industrial advantage.

Operationalizing the definition means testing deployable performance instead of inferring risk from training inputs, paired with evaluation methods that make the definition governable. The framework in this paper suggests three concrete commitments for policymakers. First, define weaponization in terms of intent and demonstrated capability rather than training compute alone. Second, invest in an independent evaluation infrastructure that can run continuous, adversarial tests across high-risk task categories, and that treats benchmark design as a validity and governance problem. Third, treat compute thresholds as downstream instruments that should be updated based on measured transition dynamics across the AI triad, not as the primary definition of what is dangerous.

This approach does not eliminate political conflict over security priorities, but it changes what must be argued about. Grounding governance in demonstrated capability and real-world use cases offer the only pathway to overcoming the LLM Mirage that has held U.S. policy-making hostage over the past five years.

\begin{acks}
Authors, as part of their affiliation with the University of California, Berkeley, were supported in part by the National Science Foundation, U.S. Department of Defense, Founders Pledge Fund, the House Fund, and the Berkeley Artificial Intelligence Research (BAIR) industrial alliance program.
\end{acks}
\clearpage
\section*{Endmatter}
\subsection*{Positionality}
We write as researchers based in the United States, working at the intersection of AI research, AI policy, and national security. This vantage point shapes our emphasis on U.S. regulatory institutions and great-power competition, and it risks reproducing a state-centric view of security that can obscure whose safety is prioritized. We therefore treat national security claims as contested political arguments rather than neutral descriptions of harm, and we foreground which communities face near-term exposure to AI-enabled violence and surveillance, and which communities benefit from compute-centric controls. Our framework is intended to support accountability and harm reduction, not to legitimize indefinite frontier advantage as an end in itself.
\subsection*{Generative AI Usage Statement}
The authors used ChatGPT 5.2 for formatting and grammar checks, and to ensure that the submission met the ACM formatting guidelines. All outputs were reviewed by the authors before making edits to the manuscript.
\bibliographystyle{ACM-Reference-Format}
\bibliography{reference}

\end{document}